\documentclass[11pt,DIV=14,abstract=true]{scrartcl}
\usepackage{graphicx}
\usepackage[dvipsnames]{xcolor}
\usepackage{amsmath,amsfonts,amssymb}
\usepackage{booktabs}
\usepackage{physics}
\usepackage{subcaption}
\usepackage{float}
\usepackage[section]{placeins}
\usepackage[utf8]{inputenc}
\usepackage{array}
\usepackage{tikz,lipsum,lmodern}
\usepackage[most]{tcolorbox}
\usepackage[backend=biber,
            style=numeric-comp,
            sorting=none,
            doi=false,
            isbn=false,
            url=false,
            maxbibnames=20,
            maxcitenames=2]{biblatex}
\usepackage{enumitem}
\usepackage{tikz}
\usepackage{amsthm}
\numberwithin{equation}{section}

\theoremstyle{definition}

\usepackage[colorlinks=true,linkcolor=blue!45!black,citecolor=blue!45!black, urlcolor=blue!45!black]{hyperref}
\bibliography{refs}

\setcapindent{0pt}
\usepackage{authblk}

\title{Complex Quantum Dynamics Versus Classical Simulability of Noisy Random Circuits}
\author{Anjali Waghmare\thanks{\texttt{anjali.waghmare7005@gmail.com}}}
\author{Sergii Strelchuk\thanks{\texttt{Sergii.Strelchuk@cs.ox.ac.uk}}}
\author{Sathyawageeswar Subramanian\thanks{\texttt{Sathya.Subramanian@cs.ox.ac.uk}}}
\affil{\small \textit{Department of Computer Science, University of Oxford, Parks Rd, Oxford OX1 3QG, United Kingdom}}
\date{}

\begin{document}
\maketitle
\begin{abstract}
 Claims of quantum advantage rest on the classical hardness of simulating quantum circuits. Magic, operator scrambling, anticoncentration, and non-Gaussianity for fermionic circuits are standard diagnostics of complex quantum dynamics. For pure states, some of these have been rigorously connected to classical simulability. Whether these diagnostics reliably track the limits of efficient classical simulation under noise remains unclear. Here, we show that in noisy Clifford+$T$ and nearest-neighbour matchgate+SWAP circuits, dynamical diagnostics and classical simulability can separate in both directions. In particular, the diagnostics can remain nontrivial after classical simulation becomes efficient, or become trivial before known efficient classical algorithms apply. We trace this mismatch to the different statistical properties they probe: magic and scrambling depend on fourth-order statistics of the Pauli spectrum, whereas the simulation algorithms depend primarily on second-order moments, which local noise suppresses at different rates. Thus, dynamical diagnostics measured on a noisy quantum device do not by themselves constitute evidence for classical hardness.
\end{abstract}
\tableofcontents

\section{Introduction}
Complexity-theoretic arguments for quantum advantage in random circuit sampling typically establish classical hardness for idealised quantum circuits under assumptions on their output distributions~\cite{Bremner2016,AaronsonChen2017,Bouland2019}. Notably, their extension to realistic noisy circuits is limited, with hardness known only for specific noise models and parameter regimes~\cite{Bouland2019,Bouland2021,Ghosh2024}. Indeed, efficient classical simulation is possible in certain noisy regimes: random circuits subject to sufficiently strong depolarising noise at every qubit and layer admit polynomial-time classical sampling algorithms~\cite{Aharonov_2023,schuster}. Thus, shallow noisy circuits implemented on present-day hardware lie in a regime where neither classical hardness nor efficient classical simulation is generally established. Since experimental quantum advantage demonstrations rely on noisy implementations, additional evidence is needed to connect experimentally observed dynamics to classical simulation complexity.

The evidence used to support such claims comes from two complementary approaches. The first is an algorithmic approach, wherein one analyses the best known classical simulation algorithms, including tensor-network contraction~\cite{pan, noh, MarkovShi2008, Gray2021hyperoptimized, Huang2020ptj}, stabilizer-rank decomposition~\cite{PhysRevX.6.021043, PhysRevLett.116.250501,Bravyi_2019, Qassim2021improvedupperbounds}, Pauli-path truncation~\cite{Aharonov_2023,schuster,angrisani2025} and degree truncation for fermionic circuits~\cite{miller2025majorana}. For each algorithm, the regime in which its computational cost remains efficient can be determined. Taken together, these algorithms delineate the region where efficient classical simulation is established; outside this region, existing methods provide no guarantee of efficient simulation.

The second approach is a dynamical one, examining the structural properties of the circuit that are expected to accompany hardness of classical simulation. The standard properties include magic (or nonstabilizerness), measured by quantities such as the stabilizer R\'enyi entropy~\cite{PhysRevLett.128.050402, haugpiroli}; operator scrambling, quantified by averaged out-of-time-order correlators~\cite{mss,hosur,nahum}, which have been measured on quantum processors~\cite{Mi2021gdf} and also used as the basis of a claim of quantum advantage~\cite{Abanin2025}; and anticoncentration of the output distribution, quantified by the inverse participation ratio~\cite{PRXQuantum.3.010333, Hangleiter2018anticoncentration}. Unlike the algorithmic criteria above, these dynamical diagnostics are defined directly from the quantum evolution, without reference to a classical simulation method.

For noiseless circuits, some such diagnostics have been rigorously connected to classical simulability. A pure state with vanishing stabilizer Rényi entropy is a stabilizer state, for which the Gottesman–Knill theorem provides an efficient classical simulation \cite{AaronsonGottesman2004}. Likewise, a pure state with little entanglement across every cut admits a compact matrix-product representation. Thus, in suitable pure-state settings, dynamical quantities can provide sufficient conditions for efficient classical simulation~\cite{Vidal2003, JozsaLinden2003}.

Real devices, however, operate under noise, and the relationship between dynamical properties and classical simulability becomes less direct once the output state is mixed. This motivates the central question of this work:

\begin{quote}
\begin{center}
\itshape
In noisy quantum circuits, do dynamical diagnostics such as magic,
scrambling, and anticoncentration reliably track classical simulability?
\end{center}
\end{quote}

We show, in two distinct random circuit families, that these diagnostics need not correlate well with classical simulability under noise. The mismatch occurs in both directions: for example, a circuit can retain nontrivial magic while already admitting efficient classical sampling, or lose its magic signature while remaining outside all regimes covered by known efficient algorithms. The separation arises because the diagnostics and the simulation algorithms probe different properties of the noisy state, which respond differently to noise.

\subsection{Summary of Results}

We study whether dynamical diagnostics remain reliable indicators of classical simulation complexity in noisy quantum circuits by studying two circuit families. The first is brickwork Clifford circuits doped with $T$ gates under local depolarising noise, where mixed-state magic and the leading classical simulation algorithms can be evaluated on the same architecture. The second is matchgate circuits doped with SWAP gates under Majorana dephasing. Since matchgate circuits are polynomial-time simulable~\cite{valiant2002,terhal2002,knill2001} and the addition of SWAP gates makes them universal~\cite{jozsamiyake,brodgalvao}, this family provides an independent setting in which any mismatch between dynamical diagnostics and simulation cost cannot be attributed to missing algorithms.

\begin{figure}[hbt]
\centering
\includegraphics[width=\textwidth]{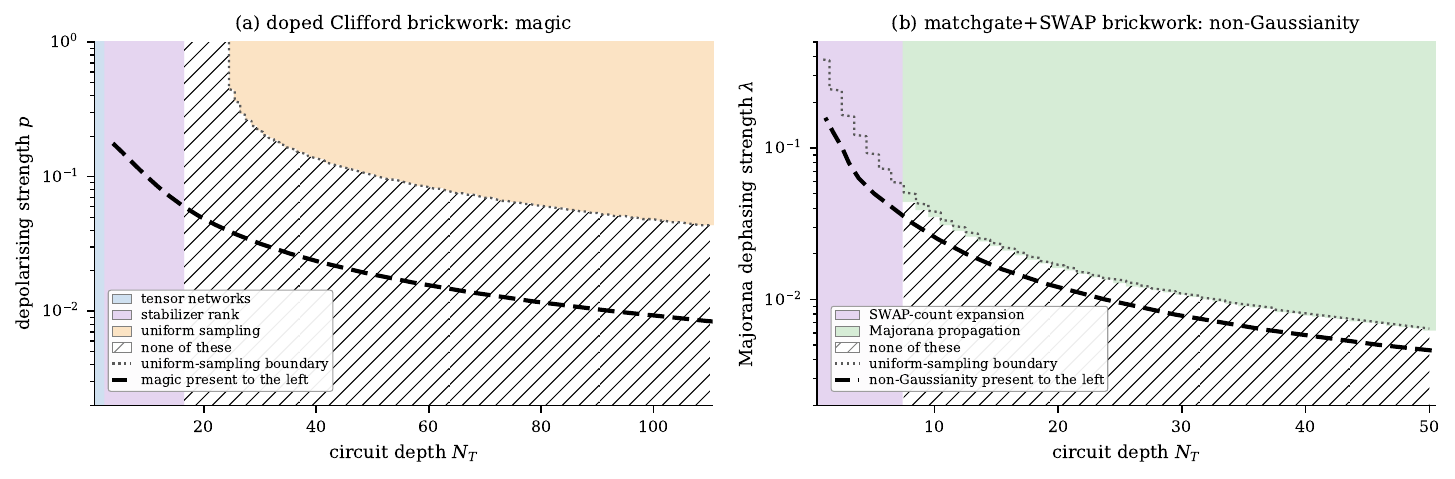}
\caption{Dynamical diagnostics and classical simulation boundaries do not coincide in noisy circuits. Each cell is coloured by the most efficient of the considered classical algorithms, the hatched region is covered by none of them, and the dotted line marks the uniform-sampling threshold. Dashed curves are the resource boundaries. (a) Brickwork Clifford$+T$ circuits with $N=6$, $\nu=2$ and one noisy qubit; the resource is mixed-state magic, quantified by $\mathcal{W}>0$ in Eq.~\eqref{eq:witness}. For sufficiently strong noise, the witness is non-positive at every depth. (b) Matchgate$+$SWAP circuits with $N=6$ and $n_S=1$ under Majorana dephasing; the resource is fermionic non-Gaussianity, Eq.~\eqref{eq:nongaussianity}. In both circuit families, the resource boundary cuts across the simulation landscape. The resource can thus remain detectable where efficient classical simulation is already possible, and disappear where no efficient simulation algorithm is known.}
\label{fig:extabstractfigure}
\end{figure}

Fig.~\ref{fig:extabstractfigure} summarises our central result. For both circuit families, the boundary beyond which the diagnostic for the resource enabling universality becomes trivial by the corresponding dynamical diagnostic does not coincide with the boundary separating regions of provably efficient classical simulation. Instead, the diagnostic boundary cuts across it: the resource can remain detectable where efficient simulation is already possible, while becoming trivial in regions where no efficient simulation algorithm is known.

Our main contributions are as follows:

\begin{enumerate}
    \item We show that magic, fermionic non-Gaussianity, scrambling, and anticoncentration do not reliably track classical simulation cost in noisy mixed-state circuits. Across both doped Clifford and doped matchgate architectures, these diagnostics can remain nonzero in regimes where efficient classical simulation is already possible, and disappear in regimes where no efficient simulation algorithm is known.

    \item We identify the mechanism underlying this mismatch. We show that local depolarising noise suppresses the $2m$-th moment of the Pauli spectrum asymptotically $m$ times faster than the second moment. Since dynamical diagnostics probe fourth moments while the leading simulation algorithms depend on second moments, the two evolve under different timescales in the presence of noise. Under sparse noise, the depth at which magic disappears grows as $\Theta(N)$, while the depth at which efficient sampling becomes possible remains $\Theta(1)$. Their intersection defines a crossover system size,
\begin{equation}
N^* = 3\log_2\!\left(\frac{1}{\varepsilon^2\delta_c}\right) + O(1),
\label{eq:Nstar}
\end{equation} where $\varepsilon$ is the error in sampling as measured by total-variation distance and $\delta_c$ is the failure probability over the random choice of circuit. The crossover depends only on the sampling tolerance and is independent of the noise strength.

\item We derive exact analytical results for noisy Clifford and matchgate dynamics. For Clifford circuits, we obtain closed-form expressions for the purity and inverse participation ratio at arbitrary depth, doping, and noise. For matchgate circuits, we show that classical simulation cost is determined by the degree distribution instead of a scalar dynamical diagnostic. We  provide a new perspective on the relationship between dynamical diagnostics and classical simulation by showing that exact fourth-order circuit averages can be computed through a transfer-network representation over the fourth-order Clifford commutant.

\end{enumerate}

\subsection{Related Works} The use of dynamical properties as indicators of classical simulation complexity is motivated by settings in which such properties are directly connected to classical simulability. Entanglement provides the clearest example: states with sufficiently limited entanglement admit efficient tensor-network descriptions \cite{Vidal2003}; more generally, multipartite entanglement is closely tied to the possibility of exponential quantum speedup in pure-state computation \cite{JozsaLinden2003}. Measurement-induced transitions between low- and high-entanglement phases have been interpreted as transitions in the efficiency of classical simulation ~\cite{PhysRevX.9.031009, PhysRevB.100.134306, Fisher_2023}.  Magic provides a complementary notion of computational complexity beyond entanglement, with quasiprobability and stabilizer-based measures providing quantitative bounds on the cost of classical simulation ~\cite{Veitch2014, Pashayan2015, PhysRevLett.116.250501}, and recent work showing dynamical transitions in magic and relating them to classical simulation in monitored Clifford+$T$ circuits~\cite{Bejan2024}. Together, these results indicate that, in suitable pure-state settings, dynamical resource measures can reflect the resources required for classical simulation.

Noise changes this picture in the following ways. First, the output state becomes mixed, so the pure-state correspondences between dynamical resources and efficient classical descriptions no longer apply directly. Second, noise itself can make quantum circuits easier to simulate. A range of classical methods exploit this effect, including tensor-network, stabilizer-based, and Pauli-expansion approaches, and have established efficient sampling or simulation in specific noisy regimes~\cite{noh,Aharonov_2023,schuster,nelson2026polynomialtimeclassicalsimulationnoisy}. Recent developments have further broadened Pauli-path and Pauli-propagation methods, extending efficient classical simulation beyond average-case random circuits and to arbitrary incoherent local noise, including nonunital channels~\cite{Gonz_lez_Garc_a_2025,angrisani2025}. Moreover, the quantities controlling the performance of these algorithms are generally distinct from the dynamical diagnostics used to characterise the circuit. Recent work has further shown that nonunital noise can qualitatively alter the output statistics of random circuits, preventing anticoncentration and thereby undermining both hardness and easiness arguments based on this property~\cite{Ghosh2024}. Together, these developments complicate the interpretation of dynamical diagnostics as evidence for classical simulation hardness.

Recent developments in Pauli-spectrum methods and higher-moment techniques provide the tools needed to analyse noisy mixed-state dynamics, including studies of noise-induced transitions in Pauli truncation~\cite{Dowling2026} and replica/Clifford-commutant methods for exact higher-moment calculations in random and doped Clifford circuits~\cite{grossnezamiwalter2021,bittel2025completetheorycliffordcommutant,magni2025quantumcomplexitychaosmanyqudit}. A closely related work establishes noise-dependent conditions for efficient classical simulation when noise is confined to the injected magic resource, while the underlying Clifford or matchgate dynamics remain ideal~\cite{Heo2026}. What remains unclear is whether quantities such as magic, scrambling, and anticoncentration continue to provide reliable evidence about classical simulability once noise acts throughout the circuit, and the output state is mixed. We address this question for two circuit families and show that these diagnostics need not reliably track classical simulation cost.

\paragraph{Organisation.} The remainder of this paper is arranged as follows. Sec.~\ref{sec:setting} introduces the circuit families, dynamical diagnostics, and classical simulation algorithms that define our setting. In Sec.~\ref{sec:mechanism} the theoretical mechanism underlying the separation between dynamical and algorithmic evidence in noisy circuits is developed. Sec.~\ref{sec:framework} presents the exact fourth-moment computations. Secs.~\ref{sec:resultscliffordcircuits} and \ref{sec:resultsmatchgatecircuits} apply this framework to doped Clifford and matchgate circuits respectively, showing the lack of correspondence between the dynamical diagnostics and classical simulability for these two settings. Finally, Sec.~\ref{sec:discussion&outlook} discusses the implications for experimental demonstrations of quantum advantage and outlines open directions.   

\section{Preliminaries and Setting}
\label{sec:setting}

We compare dynamical diagnostics and classical simulation algorithms on two  families of noisy quantum circuits. Throughout, we consider $N$ qubits with Hilbert-space dimension $D=2^N$, initial state $|0\rangle^{\otimes N}$, and circuit depth $N_T$.

\subsection{Circuit Families and Noise}

\begin{figure*}[t]
\centering
\includegraphics[width=0.65\textwidth]{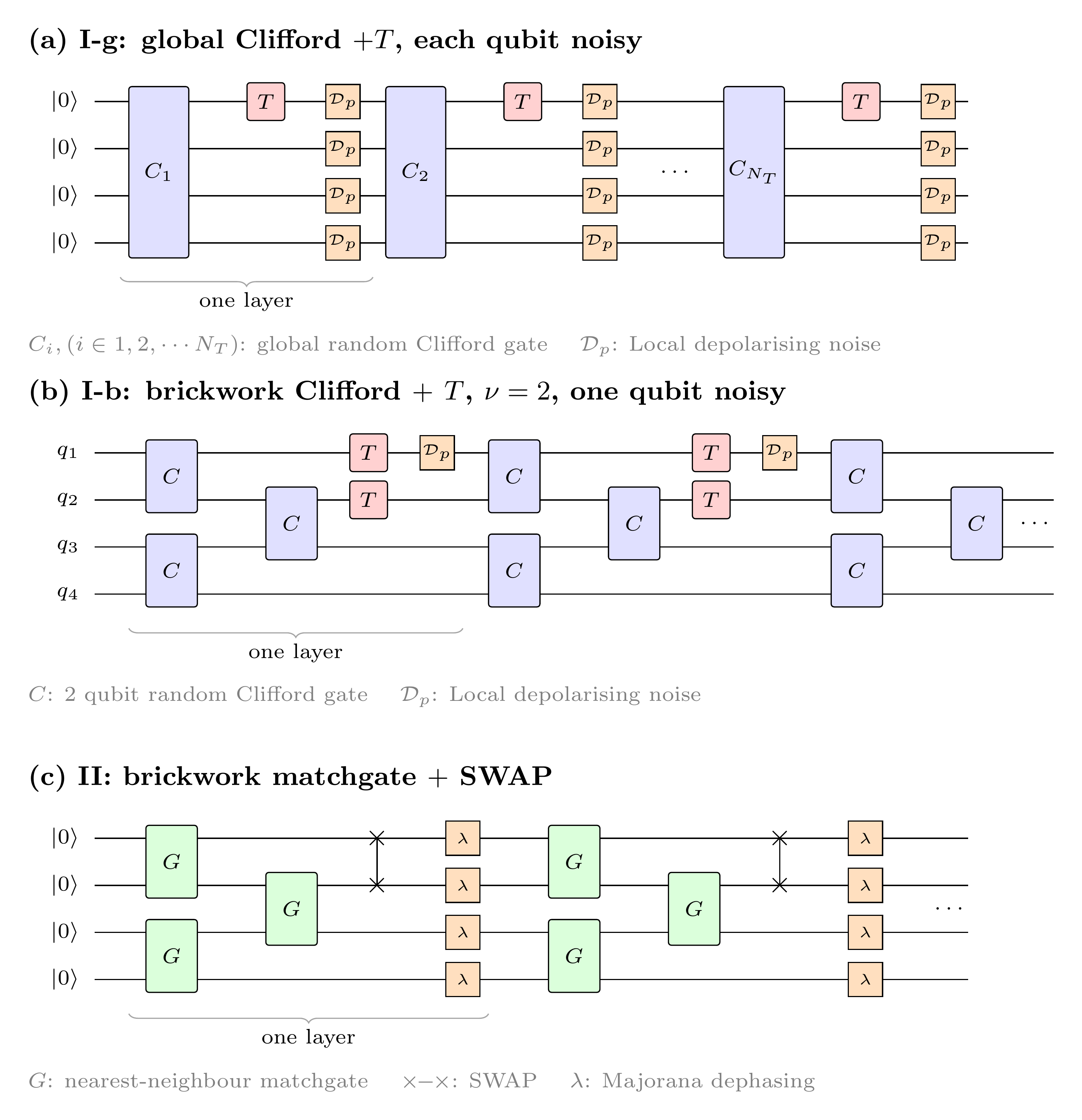}
\caption{Circuit architectures. (a) I-g: a Haar-random Clifford on all qubits, then $\nu$ $T$ gates each preceded by its own Clifford, then depolarising noise of strength $p$. (b) I-b: alternating layers of random two-site Clifford gates, $\nu$ $T$ gates per layer, and noise restricted to a single qubit. (c) II: a matchgate brickwork, $n_S$ nearest-neighbour SWAP gates per layer, and Majorana dephasing of strength $\lambda$ on every mode.}
\label{fig:architectures}
\end{figure*}

We study two circuit families, one based on Clifford dynamics and the other on matchgate dynamics, each with a distinct framework for efficient classical simulation. The architectures based on these two families are illustrated in Fig.~\ref{fig:architectures}.

\paragraph{Doped Clifford circuits.}
For each layer, we apply an entangling step, a doping step and a noise step. In the global model (I-g), as shown in Fig.~\ref{fig:architectures} (a), the entangling step is a uniformly random element of the Clifford group $\mathrm{Cl}_N$. For the brickwork model (I-b), as shown in Fig.~\ref{fig:architectures} (b), the entangling step is a layer of independent uniformly random two-qubit Clifford gates  arranged in a nearest-neighbour brickwork pattern. For both architectures, this is followed by $\nu$ single-qubit $T$ gates applied to the qubits, and finally a local noise channel. We consider single-qubit depolarising noise of strength $p$,
\begin{equation*}
    \mathcal{D}^{(j)}_p(\rho)
=(1-p)\rho+\frac{p}{2}\,I_j\otimes\mathrm{Tr}_j(\rho)
\end{equation*}
applied independently to each qubit in a noisy set $\mathcal{M}$ after every circuit layer. Its action on Pauli operators $P$ is diagonal,
\begin{equation*}
    \mathrm{Tr}\!\left[P\,\mathcal{D}^{(j)}_p(\rho)\right] = \begin{cases}
    r\,\mathrm{Tr}(P\rho), & P \text{ acts nontrivially on } j,\\
    \mathrm{Tr}(P\rho), & \text{otherwise},
\end{cases}
\end{equation*}

where $r = 1-p$.

\paragraph{Doped matchgate circuits.}
For each layer, we apply two qubit matchgates arranged in a nearest-neighbour brickwork pattern, followed by $n_S$ nearest-neighbour SWAP gates and Majorana dephasing. Matchgate circuits are polynomial-time classically simulable, whereas the addition of SWAP gates renders the architecture universal. Under the Jordan--Wigner transformation, the $N$ qubits are represented by Majorana operators $\gamma_1,\ldots,\gamma_{2N}$, and Pauli strings correspond to Majorana monomials $\Gamma_S=\prod_{j\in S}\gamma_j$.
Majorana dephasing acts diagonally on the monomials,

\begin{equation*}
    \Gamma_S \longrightarrow \tilde r^{\,|S|}\Gamma_S,
\qquad
\tilde r=1-2\lambda
\end{equation*}
preserving the fermionic structure of the dynamics.

We consider two noise geometries. One is the dense-noise regime, where depolarising noise acts on every qubit after each layer, $\mathcal{M}=\{1,\ldots,N\}$. The other is the sparse-noise regime, where the number of noisy qubits remains fixed as $N$ increases. As we show later, these two geometries lead to qualitatively different relationships between dynamical diagnostics and classical simulation. Under dense noise, Pauli path truncation is efficient at arbitrary depth, so the entire depth–noise plane is eventually covered by provably efficient classical simulation. Under sparse noise, however, such a region survives, making it possible to test whether dynamical diagnostics provide additional evidence for classical simulability. 
 
Our choice of architectures is motivated by the requirements in which regions not ruled out by any known efficient simulation algorithm exist at all. The doping density must be large enough that the stabilizer-rank boundary lies below the depth at which the output distribution becomes indistinguishable from the uniform, which requires 
\begin{equation}
  \nu\gtrsim \frac{c\log_2N}{(\alpha N_T^{\mathrm {unif}})}
\end{equation}
and is met by $\nu = 2$ for the sizes we compute. The noise must be sparse. In the case of the global Clifford architecture, anticoncentration sets in at the first layer itself; thus the circuit is classically simulable very early on, making this case less interesting than the local models where a region not ruled out by classical simulation algorithms persists. Therefore the simulability maps are drawn for the local circuits only. 

\subsection{Pauli Representation and Dynamical Diagnostics}

Any $N$-qubit mixed state can be expanded in the Pauli basis as
\begin{equation}
\rho=\frac{1}{D}\sum_{P\in\mathcal{P}_N} a_P\,P,
\qquad
a_P=\Tr(P\rho),
\label{eq:pauli_expansion}
\end{equation}
where $\mathcal{P}_N$ denotes the $N$-qubit Pauli group. The set of coefficients $\{a_P\}$ constitutes the Pauli spectrum of the state. Since the depolarising channel acts diagonally on these coefficients, both the dynamical diagnostics and the classical simulation criteria considered in this work can be expressed in terms of moments of the Pauli spectrum.

We consider four dynamical diagnostics: magic, scrambling, anticoncentration, and fermionic non-Gaussianity. 

\paragraph{Anticoncentration.}
Let $p(x)=\langle x|\rho|x\rangle$ denote the computational-basis output distribution. The normalised 2nd order inverse participation ratio (equivalently, the collision probability) is
\begin{equation}
\mathcal{A}
= D\sum_x p(x)^2
= \sum_{P\,\mathrm{diagonal}} a_P^2,
\label{eq:anticonc}
\end{equation}
where the second equality follows from the Fourier expansion of computational-basis projectors in the diagonal Pauli basis. A uniform distribution satisfies $\mathcal{A}=1$, while a Haar-random pure state has
$\mathcal{A}=2D/(D+1)$, the Porter--Thomas value. We therefore regard a distribution as anticoncentrated when $\mathcal{A}=O(1)$~\cite{PRXQuantum.3.010333,Magni_2025}.

\paragraph{Magic.}
We characterise the quantum state using purity and stabilizer purity

\begin{equation}
\Delta_2=\frac1D\sum_Pa_P^2=\mathrm{Tr}\rho^2,\qquad \Delta_4=\frac1D\sum_Pa_P^4 ,
\label{eq:purities}
\end{equation}

The fourth moment $\Delta_4$ is the expectation of the $\Omega_4 = D^{-1} \sum_P P^{\otimes 4}$ in $\rho^{\otimes 4}$ \cite{bittel2025completetheorycliffordcommutant}, which makes it directly computable within the fourth-order Clifford commutant.

For pure states, the Stabilizer R\'enyi Entropy is $M_2 = -\text{log} \Delta_4$, it vanishes exactly on stabilizer states and attains the Haar-typical value $\text{log}_2 (D+3)/4$. For mixed states, however, $M_2$ alone is not a faithful measure of magic \cite{Haug_2026}. A small $M_2$ can result either from non-stabilizerness or from classical mixing. We therefore  use the purity-normalized magic witness 
\begin{equation}
\mathcal{W}\equiv M_2-2S_2
=3\log_2\Delta_2-\log_2\Delta_4,
\label{eq:witness}
\end{equation}
where $S_2=-\log_2\tr(\rho^2)$ is the $2$-Rényi entropy. Equivalently,
$\mathcal{W}>0$ iff $\Delta_2^3>\Delta_4$. On pure states, $\mathcal{W}$ reduces to the Stabilizer R\'enyi Entropy $M_2$. 

\paragraph{Scrambling.}
We quantify scrambling using the averaged out-of-time-order correlator (OTOC), which probes operator spreading under Heisenberg evolution~\cite{mss,hosur,nahum}. Given two Pauli operators $A$ and $B$, the  four-point OTOC is
 \begin{equation}
\mathrm{OTOC}(t)=\frac{1}{D}\mathrm{Tr}\!\left[A(t)BA(t)B\right],
\label{eq:otoc}
\end{equation}
where $A(t)=U^\dagger(t)AU(t)$ is the Heisenberg-evolved operator.

We average Eq.~\eqref{eq:otoc} over pairs of local Pauli operators and over circuit realizations. As initially local operators spread across the system, the averaged OTOC decays from its initial value towards the Haar-random value, providing a dynamical measure of scrambling.

\paragraph{Fermionic non-Gaussianity.}
For the matchgate family, the relevant dynamical resource is fermionic non-Gaussianity. Under the Jordan--Wigner transformation, an $N$-qubit state is described by $2N$ Majorana operators $\gamma_1,\ldots,\gamma_{2N}$ satisfying $\{\gamma_i,\gamma_j\}=2\delta_{ij}$. Its covariance matrix is
\begin{equation}
\Gamma_{ij}
=\frac{i}{2}\mathrm{Tr}\!\left(\rho[\gamma_i,\gamma_j]\right),
\end{equation}
which completely characterizes fermionic Gaussian states.

Given an arbitrary state $\rho$, let $\rho_G$ denote the unique Gaussian state with the same covariance matrix. We quantify fermionic non-Gaussianity by the relative entropy
\begin{equation}
\delta(\rho)
=S(\rho_G)-S(\rho),
\label{eq:nongaussianity}
\end{equation}
where $S(\rho)=-\mathrm{Tr}(\rho\log\rho)$ is the von Neumann entropy. The measure satisfies $\delta(\rho)\ge0$ and vanishes if and only if $\rho$ is Gaussian, and is the natural analogue of mixed-state magic for the matchgate architecture.

\subsection{Classical Simulation Algorithms}
\label{sec:algorithms}

We first specify the efficiency criterion against which the classical simulation algorithms are assessed. An algorithm is considered efficient if its runtime scales polynomially with the system size, i.e. it scales as $N^{O(1)}$. A region of parameter space is said to be covered if at least one algorithm is efficient there. Direct simulation of an $N$-qubit mixed state requires $\Theta(d^{2N})$ resources, providing the exponential baseline against which efficient algorithms are compared. Table~\ref{tab:algorithms} lists the algorithms and their costs and the efficiency conditions.

\paragraph{Tensor-networks.} 

Tensor-network methods are efficient for sufficiently shallow circuits, with computational cost determined primarily by the causal light cone and the resulting bond dimension~\cite{pan,noh}.
For a brickwork circuit of depth $N_T$, tensor-network contraction has bond dimension bounded by the causal light cone,
\begin{equation}
    \chi \le d^{\min(N_T,N/2)},
\end{equation}
and contraction cost $O(NN_T\chi^3)$. The algorithm is therefore efficient when
\begin{equation}
N_T \le \frac{c\log_2 N}{3\log_2 d},
\label{eq:tnbound}
\end{equation}
up to polynomial prefactors. Stronger bounds exist for noisy circuits, where decoherence further suppresses entanglement~\cite{noh}, and we use Eq.~\eqref{eq:tnbound} as a noise-independent baseline. 

\paragraph{Stabilizer-rank simulation.}

Stabilizer decompositions provide efficient simulation when the number of non-Clifford gates grows at most logarithmically with system size~\cite{PhysRevLett.116.250501,Bravyi_2019}. The corresponding cost depends on the injected non-Clifford resource rather than on the noisy output state itself.
Stabilizer-rank simulation decomposes the circuit into a superposition of stabilizer states. For a circuit containing $t$ non-Clifford gates, the computational cost scales as
\[
\mathrm{poly}(N)\,2^{\alpha t},
\]
where $\alpha=\log_2[4/(2+\sqrt2)]\approx0.23$, the exponent characterising approximate simulation via sparsification of the stabilizer decomposition~\cite{Bravyi_2019}. The algorithm is therefore efficient whenever
\begin{equation}
t\le \frac{c\log_2 N}{\alpha},
\label{eq:stabrank}
\end{equation}
up to polynomial prefactors. Like tensor-network contraction, its cost is determined entirely by the circuit structure, in this case the non-Clifford gate count, and is independent of the noise model and the output state.

\paragraph{Pauli-path truncation.}

Pauli-path truncation approximates the output distribution by expanding it over trajectories of Pauli strings through the circuit~\cite{Aharonov_2023,schuster}. A path
$s=(s_0,\ldots,s_{N_T})$ assigns a Pauli string to each circuit layer, and its weight is
\[
|s|=\sum_{\ell}|s_\ell|.
\]
The algorithm retains only paths of weight at most $\ell$, with the resulting mean-square error bounded by
\begin{equation}
D^2\,\mathbb{E}_x\!\left[(p_\ell(x)-p(x))^2\right]
\le
R(\ell)
\equiv
\sum_{|s|>\ell}\widetilde W_s,
\label{eq:pauli_path}
\end{equation}
where $\widetilde W_s$ is the normalised weight of path $s$. Since at most
$\binom{NN_T}{\ell}3^\ell$ paths have weight $\ell$, the runtime scales as
\[
N^{O(\ell^*)},
\]
where
\[
\ell^*=\min\{\ell:R(\ell)\le\varepsilon^2\delta_c\}
\]
is the smallest truncation order achieving the target sampling accuracy. Eq.~\eqref{eq:pauli_path} requires two properties of the ensemble. The first is gate-set orthogonality: distinct Pauli paths must be uncorrelated in expectation, so that the squared error is the sum of the discarded weights. This holds whenever the gate distribution is invariant under right multiplication by a random Pauli, which is the case for uniformly random Clifford gates, global or two-local, and remains the case when fixed $T$ gates are inserted between them. We discuss this further in Appendix~\ref{app:orthogonality}. The second is anticoncentration of the ideal circuit, as a result of which normalised path weights satisfy
$\sum_s\widetilde W_s=\mathcal{A}$. The truncation error is therefore measured relative to the IPR, and this condition is the link between the algorithm and the diagnostic.

Under dense depolarising noise, every path contributing to the non-uniform part of the output is nontrivial at each of the $N_T+1$ slices, since a gate maps the identity string to itself, and is damped by at least $q^{|s|}$ with $q=r^2$. Hence
\begin{equation}
R(\ell)\le q^{\ell}(\mathcal{A}_{\mathrm {ideal}}-1),\qquad
\ell^*\le\frac{\ln[(\mathcal{A}_{\mathrm {ideal}}-1)/\varepsilon^2\delta_c]}{\ln(1/q)},
\label{eq:ellstar}
\end{equation}
independent of $N$. The algorithm is therefore polynomial-time for any fixed noise strength $p>0$.  The polynomial exponent, however, can be large, for example,
$\ell^*\approx67$ at $p=0.05$ and
$\varepsilon^2\delta_c=10^{-3}$.

\paragraph{Uniform Sampling.} If $\mathcal{A}-1\le\varepsilon^2\delta_c$, the output distribution is $\varepsilon$ close to the uniform, and can therefore be sampled trivially. 

\paragraph{Fermionic degree truncation.}

For matchgate circuits, Majorana-propagation algorithms approximate the output state by truncating its expansion in fermionic degree~\cite{miller2025majorana}. Writing the state as
\[
\rho=D^{-1}\sum_S a_S\Gamma_S,
\]
only the $\binom{N}{k}$ Majorana monomials of degree $2k$ that are diagonal in the Fock basis contribute to the output probabilities. Since their characters are exactly orthogonal, truncation at degree $2k^*$ gives
\begin{equation}
D^2\,\mathbb{E}_x\bigl[(p_{k^*}(x)-p(x))^2\bigr]=\sum_{k>k^*}\frac{\binom Nk\,W_{2k}}{\binom{2N}{2k}},
\label{eq:degree_tail}
\end{equation} where $W_{2k}$ is the total weight in the degree-$2k$ sector. 
The truncation error is the contribution from sectors with degree greater than $2k^*$, without any assumptions about the circuit ensemble. Retaining monomials up to degree $2k^*$ gives a computational cost of
\begin{equation}
    \mathrm{poly}(N)\binom{2N}{2k^*}\sim(2N)^{2k^*}
\end{equation}
This is the Majorana propagation algorithm \cite{miller2025majorana}, which provides the natural fermionic analogue of Pauli-path truncation. When the number of SWAP gates is small, the circuit can instead be expanded over the SWAP gates, at an exponential cost in their number~\cite{diaskonig,reardonsmith}. In the absence of SWAP gates, the covariance-matrix method simulates the circuit in $O(N^3)$ time~\cite{valiant2002,terhal2002,knill2001}.

\begin{table}[hbt]
  \begin{tabular*}{\textwidth}{@{\extracolsep{\fill}}lcc}
  \hline  \hline
    Algorithm & Cost & Efficiency Condition\\
    \hline
    Tensor networks~\cite{pan,noh} & $O(N N_T\chi^3)$, $\chi\le 2^{\min(N_T,N/2)}$ & $N_T\le c\log_2N/3$\\
    Stabilizer-rank~\cite{Bravyi_2019,PhysRevLett.116.250501} & $\mathrm{poly}(N)\,2^{\alpha\nu N_T}$ & $\nu N_T\le c\log_2N/\alpha$ \\ 
    Pauli-path truncation~\cite{Aharonov_2023,schuster} & $N^{O(\ell^*)}$ & $\ell^*=O(1)$ \\
    Uniform sampling & $O(N)$ & $\mathcal{A}-1\le\varepsilon^2\delta_c$\\
    \hline
    Covariance matrix~\cite{valiant2002,terhal2002,knill2001} & $O(N^3)$ & no SWAP gates\\
    SWAP-count expansion~\cite{diaskonig,reardonsmith} & $\mathrm{poly}(N)\,2^{n_S N_T}$ & $n_SN_T\le c\log_2N$ \\
    Majorana propagation~\cite{miller2025majorana} & $\mathrm{poly}(N)\binom{2N}{2k^*}$ & $2k^*\le c$ \\
    \hline \hline
    \end{tabular*}
\caption{Classical algorithms, their cost, the condition under which they are efficient. Tensor-network and stabilizer-rank methods derive their efficiency from properties of the circuit itself, like the circuit depth and the number of injected non-Clifford gates. Pauli-path truncation, uniform sampling, and fermionic degree truncation derive their efficiency from properties of the noisy output state, like the second moments of the Pauli or Majorana spectrum.}
\label{tab:algorithms}
\end{table}

\section{Mechanism}
\label{sec:mechanism}

In this section, we identify the mechanism responsible for the mismatch between dynamical diagnostics and classical simulation boundaries. We observe that local depolarising noise suppresses different moments of the Pauli spectrum at parametrically different rates. In particular, we show that the $2m$-th moment is suppressed asymptotically $m$ times faster than the second moment. As a result, the classical simulation algorithms and the dynamical diagnostics evolve on different noise and depth scales.

\subsection{Noise Damps Higher Moments Faster}

Consider one layer of the global model without doping, and let k be an even moment order. Conjugation by a Clifford permutes the non-identity Pauli strings up to a sign, and the Clifford group can map any non-identity Pauli string to any other. Averaging over the random entangling layer therefore makes every non-identity Pauli coefficient statistically equivalent, so after the entangling step every non-identity coefficient carries the same expected $k$-th power. The noise then multiplies $a_P^k$ by $r^{k|P|}$, and since the Pauli weight is additive over sites,
\begin{equation}
\sum_{P\neq\mathbb{I}}a_P^{\,k}\ \longmapsto\ \kappa_k\sum_{P\neq\mathbb{I}}a_P^{\,k},\qquad
\kappa_k=\frac{\Lambda_k^N-1}{D^2-1},
\label{eq:kappa}
\end{equation}
with $\Lambda_k=1+(d^2-1)r^k$. Writing $\kappa_k\simeq d^{-N\alpha_k}$ with $\alpha_k=\log_d(d^2/\Lambda_k)$ gives the relation,
\begin{equation}
\frac{\alpha_{2m}}{\alpha_2}\ \longrightarrow\ m \qquad (p\to0),
\label{eq:ratio}
\end{equation}
so that the fourth moment of the Pauli spectrum decays asymptotically twice as fast as the second, and higher moments faster still. Noise both removes Pauli weight and flattens its distribution, and the flattening happens twice as fast.

The argument relies on the fact that, in the global Clifford ensemble, a random Clifford maps every non-identity Pauli string to every other with equal probability after averaging. It therefore applies only to the undoped global model. The $T$ gates modify the fourth moment, and in a brickwork circuit, locality prevents this complete mixing. Under sparse noise, the surviving Pauli weight remains concentrated near the noisy site. We therefore measure the ratio of decay rates directly on the doped brickwork with the exact method of Sec.~\ref{sec:framework} and also numerically find it is close to two, with sparse and with dense noise alike (Appendix~\ref{app:numerics}).

\subsection{Algorithms Use Only Second Moments}

Among the algorithms introduced in Section \ref{sec:algorithms}, Pauli-path truncation and uniform sampling are the two Clifford simulation methods whose runtime depends on the noisy output state. Both depend on it through a second moment. Pauli-path truncation depends on the IPR through the normalisation of Eq.~\eqref{eq:pauli_path}, and uniform sampling is determined by the condition on the IPR, and Majorana propagation is controlled by Eq.~\eqref{eq:degree_tail}, which is the contribution from sectors above the truncation degree in the same second-moment decomposition. The remaining algorithms count gates and depth, not the state itself. 

\subsection{Mixed-State Magic Under Noise}

The difference in the decay rates between the second and the fourth moments does not by itself determine when a magic diagnostic becomes trivial. For a mixed-state magic measure to be meaningful, it must be normalised by the state's purity. 

Normalisation of the state implies $a_\mathbb{I} = 1$, so $D\Delta_4=\sum_Pa_P^4\ge1$ for every state. If the purity-normalised witness is positive, \(\mathcal{W}>0\), equivalently \(\Delta_2^3>\Delta_4\), then 
\begin{equation} \Delta_2^{3}>\Delta_4\ge \frac{1}{D} \qquad\Longrightarrow\qquad \operatorname{tr}(\rho^2)>D^{-1/3}, \label{eq:positivewitness} 
\end{equation}
which is equivalent to requiring the 2-R\'enyi entropy to satisfy \(S_2<N/3\). As a result, magic can only be witnessed while the state remains substantially purer than the maximally mixed state. This is independent of the circuit architecture, doping protocol, or the particular noise channel. The duration of the non-trivial magic window depends on the rate at which local depolarising noise increases the state's 2-Rényi entropy (equivalently, decreases $\mathrm{Tr}(\rho^2)$).

A single depolarising channel acting on qubit $j$ multiplies the Pauli coefficients of strings that are nontrivial on $j$ by $r=1-p$, while leaving the others unchanged. Writing
\begin{equation}
    D\Delta_2=\Sigma_0+\Sigma_1,
\end{equation}
where $\Sigma_0$ and $\Sigma_1$ are the contributions from strings that are respectively trivial and nontrivial on qubit $j$, the channel acts as
\begin{equation}
    D\Delta_2\longmapsto \Sigma_0+r^2\Sigma_1.
\end{equation}

The unaffected contribution is the purity of the reduced state,
\begin{equation}
    \Sigma_0=\frac{D}{2}\mathrm{Tr}[(\mathrm{Tr}_j\rho)^2]
\end{equation}
and the decomposition of $\rho$ in the local Pauli basis implies

\begin{equation}
     \mathrm{Tr}[(\mathrm{Tr}_j\rho)^2] \ge \frac{1}{2} \mathrm{Tr}(\rho^2)
\end{equation}
Hence $\Sigma_0\ge(\Sigma_0+\Sigma_1)/4$, where the inequality follows from the monotonicity of purity under the partial trace. Thus
\[
\Sigma_0+r^2\Sigma_1
= r^2(\Sigma_0+\Sigma_1)+(1-r^2)\Sigma_0
\ge \frac{1+3r^2}{4}(\Sigma_0+\Sigma_1),
\]
so a single noisy qubit can reduce the purity by at most the factor
$(1+3r^2)/4$.

Since unitary gates preserve purity, every realisation of a circuit with noisy set $\mathcal M$ satisfies
\begin{equation}
\mathrm{Tr}(\rho^2)\ge
\left(\frac{1+3r^2}{4}\right)^{|\mathcal M|N_T}
=d^{-\alpha_2|\mathcal M|N_T}.
\label{eq:entropy}
\end{equation}

Combining Eq.~\eqref{eq:entropy} with the condition
$\mathrm{Tr}(\rho^2)>D^{-1/3}$ from Eq.~\eqref{eq:positivewitness} shows that the witness can remain positive at least until $N/(3\alpha_2|\mathcal M|)$.
For sparse noise this scales as $\Theta(N)$, whereas for dense noise it is $\Theta(1)$. The bound is only an upper limit on the depth over which the witness remains positive, but the exact calculations of Sec.~\ref{sec:resultscliffordcircuits} show that, once the injected magic has saturated, the witness persists until close to this limit.

\subsection{The Crossover System Size}
\label{sec:crossover}

The magic threshold and the classical sampling threshold exhibit different asymptotic scaling with circuit depth. The excess inverse participation ratio, $\mathcal{A}-1$, is initially $\Theta(1)$ and decays exponentially with depth at a per-layer rate $\lambda_1(p)$ that is independent of the system size for a fixed noisy set. As a result, the depth at which the output distribution becomes indistinguishable from uniform is

\begin{equation}
N_T^{\mathrm {unif}}\simeq \frac{\ln[(\mathcal{A}_{\mathrm {ideal}}-1)/\varepsilon^2\delta_c]}
{\ln(1/\lambda_1)} =\Theta(1),
\label{eq:Nunif}
\end{equation}

where $\lambda_1(p)$ is obtained exactly from the second-moment transfer matrix in Sec.~\ref{sec:transfermatrixalg}.
The purity follows a different scaling. It begins at $\Tr(\rho^2)=1$ and must decrease to the magic threshold, $\mathrm{Tr}(\rho^2)=D^{-1/3}$. From Eq.~\eqref{eq:entropy}, this requires $\Theta(N)$ layers under sparse noise. Since both the purity and the IPR are second moments of the same noisy state, they decay with the same asymptotic rate $\lambda_1$. Equating the purity threshold with the constant-depth sampling threshold gives the crossover system size,

\begin{equation}
N^*=3\log_2\!\left(\frac{1}{\varepsilon^2\delta_c}\right)+O(1),
\end{equation}

for which the dependence on $\lambda_1$ cancels. The crossover therefore depends only on the sampling accuracy $(\varepsilon,\delta_c)$ and not on the noise strength.

This crossover determines the ordering of the two thresholds. For $N<N^*$, the purity reaches the magic threshold before the output satisfies the uniform-sampling criterion, so the witness becomes trivial while no efficient sampling algorithm is yet guaranteed. For $N>N^*$, the ordering is reversed: the output becomes indistinguishable from uniform before the purity threshold is reached, leaving a regime in which the witness remains positive despite the existence of an efficient classical sampler. Thus, the relative position of the magic and sampling thresholds is determined by the system size and the sampling tolerance. Under dense noise, both thresholds occur at $\Theta(1)$ depth, and the intermediate regime is removed by the applicability of Pauli-path truncation.

\section{Exact Evaluation of the Diagnostics and Simulation Measures}
\label{sec:framework}

We now describe the exact evaluation of the quantities shown in the simulability maps. The dynamical diagnostics are fourth-order moments of the Pauli spectrum and are computed using the fourth-order Clifford commutant, whereas the algorithmic quantities depend on second moments and Pauli-weight distributions and are evaluated using transfer-matrix techniques.

\subsection{Fourth-order Clifford Commutant}

To evaluate the fourth moments appearing in the dynamical diagnostics, we require the fourth-order Clifford commutant. Averaging a degree-$k$ polynomial in the density matrix over the circuit ensemble corresponds to the twirling operation

\begin{equation}
    X \longmapsto \mathbb{E}_U [U^{\otimes k} X U^{\dagger \otimes k}]
\end{equation}
For $k\leq 3$, the Clifford group forms a unitary 3-design, and its commutant is spanned entirely by the permutation operators $T_\sigma$. In this regime, the Haar–Weingarten calculus applies directly. At fourth order, which is the lowest order at which magic is visible, the Clifford commutant is strictly larger than the permutation algebra\cite{grossnezamiwalter2021}. 

In addition to the 24 permutation operators, the commutant contains operators generated by

\begin{equation}
    \Omega_4 = \frac{1}{D} \sum_{P \in \mathcal{P}} P^{\otimes 4}
\end{equation}
This contributes six additional basis elements, giving a commutant of dimension

\begin{equation*}
    24+6 = 30 
\end{equation*} \cite{grossnezamiwalter2021, bittel2025completetheorycliffordcommutant}. We denote this 30-dimensional basis by $\mathcal{B}$. 
The fourth-order twirl, $\Phi^{(4)}$, is the orthogonal projector onto the commutant and is therefore self-adjoint with respect to the Hilbert--Schmidt inner product. Expanding the twirled operator in the commutant basis, \begin{equation}
\Phi^{(4)}(X)=\sum_{\Omega\in\mathcal{B}} c_\Omega\,\Omega,
\end{equation}
the coefficients are obtained from the Gram matrix of the basis,
\begin{equation}
c=W^{-1}v,
\end{equation}
with
\begin{align}
W_{\Omega\Omega'} = \operatorname{Tr}\!\left(\Omega^\dagger\Omega'\right),\qquad
v_\Omega = \operatorname{Tr}\!\left(\Omega^\dagger X\right).
\end{align}

This is the Clifford analogue of the Weingarten formula \cite{bittel2025completetheorycliffordcommutant}. 

A direct implementation of this projection would act on $16^N$-dimensional operators. The simplification that makes this tractable is that the commutant factorises over qubits. After regrouping the four replicas of each qubit,
\begin{align}
T_\pi = t_\pi^{\otimes N},\quad
\Omega_4 T_\kappa = \left(\tfrac{1}{2}\,\omega\, t_\kappa\right)^{\otimes N},
\end{align}
where $t_\pi$ is the single-qubit replica permutation and
\begin{equation}
\omega=\sum_p p^{\otimes 4}
\end{equation}
is the single-qubit Pauli operator. Every overlap in the Gram matrix therefore factorises into the $N$-th power of a $16\times16$ single-site trace. The same factorisation applies to the initial state, the observables defining $\Delta_2$ and $\Delta_4$, and the diagonal projectors defining $\mathcal{A}$. As a result, for the global circuit model each complete layer is represented by a fixed $30\times30$ transfer matrix, independent of system size, acting on the 30 coefficients of $\mathbb{E}[\rho^{\otimes 4}]$.

\subsection{Label Fusion for Brickwork Circuits}
\label{sec:fusion}

The global construction applies a single Clifford twirl to the entire system. In the brickwork architecture, however, each Clifford gate acts only on a pair of neighbouring qubits, so the twirl becomes two-local. The fourth-order twirl over the two-qubit Clifford group projects onto the subspace
$\mathrm{span}\{M_\Omega\otimes M_\Omega\}$
where $M_\Omega$ denotes the single-qubit factor of the commutant basis introduced above. Since every element of the two-qubit commutant factorises in this form, the two qubits acted upon by a Clifford gate emerge with a common commutant label.

The ensemble-averaged fourth moment can therefore be represented as a tensor $c_{\vec{\Omega}}$ whose indices correspond to commutant labels. Qubits sharing the same label are grouped. The first brickwork layer partitions the chain into $N/2$ two-qubit blocks. The subsequent layer acts on neighbouring pairs, merging one qubit from each adjacent block while leaving the remaining qubits unpaired. After this update, the tensor contains $N/2+1$ independent label blocks, and this number remains fixed throughout the circuit.

Between successive Clifford gates, each qubit undergoes only single-qubit operations, the inserted $T$ gates and the depolarising channel. We collect these into a local map $S_j$, representing the accumulated single-qubit evolution on qubit $j$ since its previous Clifford gate. When a two-qubit Clifford acts on the pair $(i,j)$, the incoming labels $\Omega_i$ and $\Omega_j$ are fused into a new label $\Omega$ through the contraction
\begin{equation}
G_{\Omega;\Omega_i\Omega_j}=\sum_{\Omega'}(W_2^{+})_{\Omega\Omega'}
\times\mathrm{Tr}\!\left[M_{\Omega'}^\dagger S_i(M_{\Omega_i})\right]
\mathrm{Tr}\!\left[M_{\Omega'}^\dagger S_j(M_{\Omega_j})\right],
\label{eq:gate}
\end{equation}
where the Gram matrix is
\[
(W_2)_{\Omega\Omega'}
=\left[\mathrm{Tr}(M_\Omega^\dagger M_{\Omega'})\right]^2,
\]
and $W_2^+$ denotes its pseudoinverse. The pseudoinverse accounts for the single linear dependence among the 30 basis elements in the $N=2$ case. Once the contraction is performed, the accumulated maps $S_i$ and $S_j$ have been incorporated into the tensor and are reset to the identity.

The dynamical diagnostics are obtained by contracting the final tensor with appropriate single-site boundary vectors, $\Omega_4$ for the stabilizer purity, the replica-swap operator for the purity, and the diagonal projector for the normalised IPR.

\subsection{Closed Forms for the Global Model}
\label{sec:closedformglobal}

The global Clifford model does not possess a finite light cone, and therefore its dynamics cannot be compared directly with the tensor-network or Pauli-path simulation boundaries derived for local circuits.  It does, however, give closed forms that show how the dynamical diagnostics (mixed-state magic, the inverse participation ratio (IPR)) depend on depth, doping and noise.

The second moments are insensitive to the doping density. In the global model this follows from the unitary design property that the $k=2$ commutant is spanned by the permutation operators $T_e$ and $T_s$, both of which are invariant under $U^{\otimes 2}$ for any Clifford unitary. So, fixed $T$ gates inserted between random Clifford layers do not affect the second moment quantities. The same conclusion extends to brickwork architecture because the depolarising channel is invariant under unitary conjugation. Each fixed $T$ may therefore be commuted through the noise and combined with the neighbouring two-qubit Clifford gate, leaving the second moment unchanged.

As a result, both the purity and the IPR are independent of the doping parameter $\nu$, implying that second moments alone cannot detect magic. We thus need a fourth moment to see the effect of magic. Using the noise damping factor from Eq.~\eqref{eq:kappa} for the weight after $N_T-1$ layers, applying the final Clifford twirl and then the depolarising noise step gives

\begin{equation}
\begin{aligned}
\mathcal{A}(N_T) =1+\frac{w^N-1}{D+1}\,\kappa_2^{N_T-1},\qquad w=1+(d-1)r^2,\qquad
\Delta_2(N_T) =\frac{1+(D-1)\kappa_2^{N_T}}{D},
\end{aligned}
\label{eq:closed2}
\end{equation}
which interpolate between the Porter-Thomas value at $r=1$ and the uniform distribution for $r = 0$. 

The fourth moment does depend on the doping, and its dependence can be seen explicitly in the noiseless case. A $T$ gate acting on a single qubit leaves Pauli strings containing $\mathbb{I}$ or $Z$ unchanged, while rotating each pair of coefficients $(u,v)=(a_{XP'},a_{YP'})$ by an angle $\pi/4$. The fourth-moment contribution therefore transforms as 
\begin{equation}
u'^4+v'^4=\tfrac12(u^4+v^4)+3u^2v^2 .
\label{eq:uv}
\end{equation}

Averaging over the preceding Clifford layer and using
\begin{equation}
\sum_{\{P,Q\}=0}a_P^2a_Q^2=\tfrac{D^2}{2}\left(1-\Delta_4\right),
\label{eq:acpair}
\end{equation}
gives the decay factor 
\begin{equation}
    \lambda_-=\frac{3D^2-3D-4}{4(D^2-1)},
\end{equation} 
which is derived in Appendix~\ref{app:commutant}. The fixed point is $4/(D+3)$, corresponding to the Haar value. 

After $t$ non-Clifford gates,

\begin{equation}
\Delta_4(t) =\frac{4+(D-1)\lambda_-^{t}}{D+3}.
\label{eq:delta4t}
\end{equation}

Equation~\eqref{eq:delta4t} therefore gives the exact finite-depth evolution of the stabilizer purity. The fourth moment approaches its Haar value after $\Theta(\log D/ \log (4/3))=\Theta(N)$ non-Clifford gates, consistent with Refs.~\cite{Magni_2025,turkeshimagic,tirrito,szombathy,Haug2025}, and Eq.~\eqref{eq:delta4t} gives the finite-$t$ form.

\subsection{Transfer-Matrix Evaluation of Algorithmic Quantities}
\label{sec:transfermatrixalg}

The algorithmic boundaries are obtained from three exact transfer-matrix calculations. Together, these determine the second moments and weight distributions required by the classical simulation criteria. The explicit matrix elements are given in Appendix~\ref{app:transfer}. 

The first calculation evaluates the second moment of the brickwork circuit. On the replica basis $\{e,s\}^N$, a two-qubit Clifford gate is represented by the transfer matrix with
\begin{equation}
    T(e|ee)=T(s|ss)=1,\qquad
T(e|es)=T(s|es)=\frac{d}{d^2+1},
\label{eq:gateTM}
\end{equation}
and the depolarising noise acts through a local $2\times2$ transfer matrix on each noisy qubit. Contracting the network with the appropriate boundary vectors gives the normalised IPR, $\mathcal{A}$, and the purity, $\mathrm{Tr}(\rho^2)$. 

The second calculation determines the Pauli-path weight distribution. Introducing a parameter $z$ that weights each non-identity site in a Pauli path gives the generating function
\begin{equation}
G(z)=\sum_s z^{|s|}\widetilde W_s,
\end{equation}
with the normalisation $G(1)=\mathcal{A}$. The truncation order $\ell^*$ is obtained from the Chernoff bound,
\begin{equation}
    R(\ell)\le
\min_{z\ge1}
z^{-\ell}\!\left[G(z)-1\right]
\end{equation}
which is used to determine whether the $N$-independent scaling of Eq.~\eqref{eq:ellstar} persists under sparse noise.

Another first-moment argument provides a lower bound. Writing $T=\mathcal{A}-1$, with mean path weight $\langle|s|\rangle$ and maximum path weight $M=N(N_T+1)$, we get
\begin{equation}
    \ell^*>
\langle|s|\rangle
-\frac{\varepsilon^2\delta_c\,M}{T}.
\end{equation}
This shows that an increasing mean path weight forces a growing truncation order.

The third is the Majorana degree distribution of the matchgate family, which we discuss in Sec.~\ref{sec:resultsmatchgatecircuits}.

\section{Doped Clifford circuits}
\label{sec:resultscliffordcircuits}

We now construct the simulability map shown in Fig.~\ref{fig:extabstractfigure}(a). The boundaries are calculated for the brickwork Clifford circuit with doping density $\nu=2$ and a single noisy qubit. Each point on the map is assigned to the first algorithm in Table~\ref{tab:algorithms} that satisfies its efficiency criterion.

The uniform-sampling boundary is given by the exact depth at which $\mathcal{A}-1=\varepsilon^2\delta_c$, obtained from the second-moment transfer matrix. The magic boundary is given by the exact depth at which the witness $\mathcal{W}$ becomes zero, computed from the fourth-moment label-fusion network.

\subsection{Second and Fourth Moments of the Brickwork Circuit}

\begin{figure*}[t]
\centering
\includegraphics[width=\textwidth]{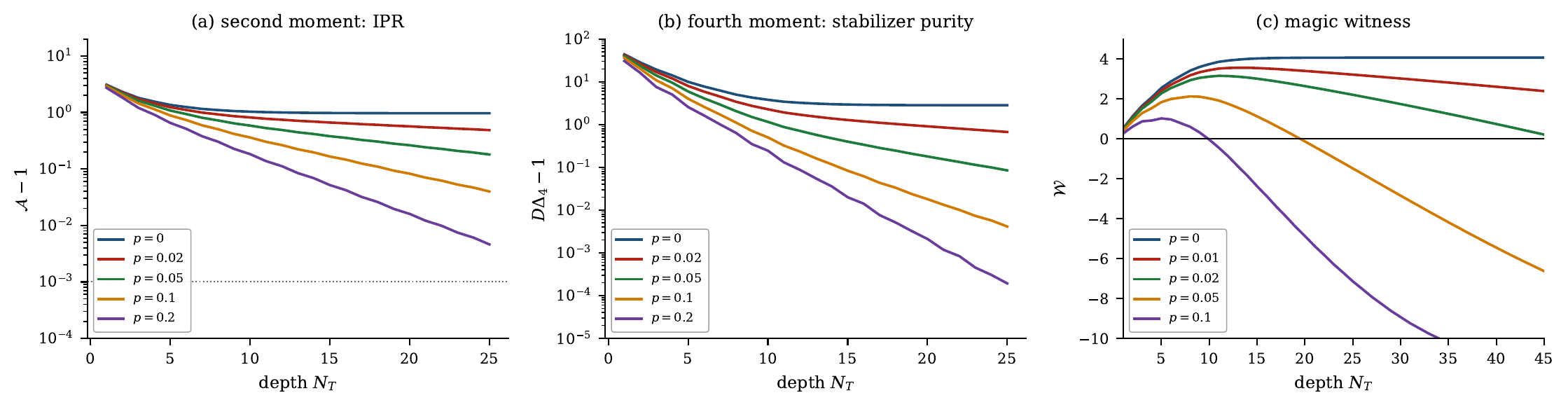}
\caption{The two moments of the brickwork circuit with $N=6$, $\nu=2$ and one noisy qubit. (a) Excess IPR $\mathcal{A}-1$; the dotted line is the sampling tolerance $\varepsilon^2\delta_c=10^{-3}$. (b) Excess stabilizer purity $D\Delta_4-1$, which after a transient decays at twice the rate of (a). (c) The magic witness $\mathcal{W}$ of Eq.~\eqref{eq:witness}, positive over a window in depth that closes as the noise grows.}
\label{fig:moments}
\end{figure*}

Fig.~\ref{fig:moments} compares the second and fourth moments for the same brickwork circuit, evaluated at identical depths. The IPR excess, $\mathcal{A}-1$, and the stabilizer purity, $\Delta_4$, exhibit the exponential decay discussed in Sec.~\ref{sec:framework}, with the fourth moment relaxing at approximately twice the rate of the second, in agreement with Eq.~\eqref{eq:ratio}.

The magic witness $\mathcal{W}$ remains positive only over a finite depth interval. At shallow depths, the injected $T$ gates generate magic more rapidly than it is removed by noise, causing $\mathcal{W}$ to increase. As the circuit becomes deeper, depolarising noise dominates the dynamics and $\mathcal{W}$ decreases to zero. Increasing the noise strength progressively reduces this interval until it disappears entirely, indicating that entropy is produced faster than magic can accumulate. The boundary defined by $\mathcal{W}=0$ therefore forms the magic threshold shown in Fig.~\ref{fig:extabstractfigure}(a). Beyond a finite noise strength, $\mathcal{W}$ is non-positive at all depths.

\subsection{Simulability Map Analysis}

At shallow depths, the stabilizer-rank algorithm is efficient, while the magic witness remains positive. This corresponds to a region in which the circuit contains non-Clifford resources but is nevertheless classically simulable because the number of $T$ gates is still small. Beyond the stabilizer-rank boundary lies an uncovered region, where none of the algorithms in Table~\ref{tab:algorithms} is provably efficient. The magic boundary passes through this region, separating circuits with $\mathcal{W}>0$ from those with $\mathcal{W}\le0$.

Fig.~\ref{fig:extabstractfigure}(a) is shown for $N=6$, which lies below the crossover size $N^*$ of Eq.~\eqref{eq:Nstar}. In this regime, the magic boundary is reached before the uniform-sampling boundary. For larger system sizes, $N>N^*$, the ordering is reversed; the uniform-sampling boundary occurs at a shallower depth, while the magic boundary extends to greater depths. 

\subsection{The Role of Noise Geometry}
\label{sec:roleofnoisegeometry}

When every qubit is noisy, the entire simulability map is covered by known classical algorithms. For depths below $N_T=c\log_2N/3$, tensor-network contraction is efficient because the causal light cone bounds the bond dimension. At slightly greater depths, the stabilizer-rank algorithm applies, and beyond this Pauli-path truncation remains efficient with a truncation order $\ell^*$ that is independent of the system size until the output approaches the uniform distribution. The generating-function calculation of Sec.~\ref{sec:transfermatrixalg} confirms this behaviour: the exact truncation order remains essentially constant in $N$ and closely matches the analytic prediction of Eq.~\eqref{eq:ellstar}.

This $N$-independent scaling relies on every surviving Pauli path being damped at each circuit layer, and this condition is satisfied under dense noise. When the noisy set is made sparse, this assumption no longer holds because paths supported away from the noisy qubits are unaffected by the depolarising channel. The first-moment lower bound of Sec.~\ref{sec:transfermatrixalg} then shows that the truncation order grows genuinely with system size. For a single noisy qubit, the mean path weight increases linearly with $N$, and both the upper and lower bounds on $\ell^*$ grow accordingly. As a result, the Pauli-path region no longer covers the full parameter space, giving rise to a region not ruled out by classical simulation algorithms.

Noise acting on a constant fraction of the qubits suppresses Pauli paths throughout the circuit and eliminates the uncovered region. Sparse noise is therefore used here as a regime in which an uncovered region exists, allowing the relationship between dynamical diagnostics and classical simulation to be examined.

\section{Doped matchgate circuits}\label{sec:resultsmatchgatecircuits}

The mismatch between dynamical diagnostics and classical simulability identified in Sec.~\ref{sec:resultscliffordcircuits} leaves open the possibility that additional classical simulation algorithms could eliminate the apparent gap. Doped matchgate circuits provide a stronger test, since nearest-neighbour matchgate circuits without SWAP gates are classically simulable in $O(N^3)$ time ~\cite{valiant2002,terhal2002,knill2001}. We show that the mismatch between dynamical diagnostics and classical simulability arises in this setting too.

\subsection{Majorana Sector Weights and Degree Evolution}\label{sec:mgsectors}

A matchgate acts on the Majoranas by an element of $SO(2N)$ and preserves $|S|$ exactly. This conservation forms the basis of matchgate simulability, and it also changes the replica structure. For Clifford ensembles, the second-moment commutant is two-dimensional,
so the second moment is fully characterised by
Eq.~\eqref{eq:gateTM}, and the asymptotic value satisfies
$\mathcal{A} \to 2$. For Gaussian ensembles, each sector weight
\begin{equation}
W_{2k}=\sum_{|S|=2k}\big(\operatorname{Tr}\rho\,\Gamma_S\big)^2
  \label{eq:Wdef}
\end{equation}
is independently conserved under the action of the matchgate layer. Thus, instead of a single conserved quantity, there are $N+1$ independent invariants, so $\mathcal{A}$ in general should not attain the Porter--Thomas value.

The initial state is Gaussian, with its Pauli spectrum supported on the $D$ products of
$Z_j=-i\gamma_{2j-1}\gamma_{2j}$. These are the diagonal strings entering Eq.~\eqref{eq:anticonc}, giving

\[
W_{2k}(0)=\binom Nk,
\qquad
\sum_{k=0}^{N}W_{2k}=D.
\]

For a global variant, this statement is exact: the signed mode permutations with unit determinant form a subgroup of $SO(2N)$, and Haar averaging makes the coefficients $\{a_S^2\}$ exchangeable within each sector, and $\mathbb{E}[a_S^2]=\dfrac{W_{2k}}{\binom{2N}{2k}}$. For a nearest-neighbour brickwork circuit, by contrast, degree equilibration is reached only after the light cone has traversed the chain, at a depth of order $N/2$ below that the output is concentrated on fewer strings and $\mathcal{A}$ is correspondingly larger. 

Majorana dephasing preserves the sector decomposition while damping sector $2k$ by a factor $\tilde r^{\,4k}$ per layer. The resulting inverse participation ratio is

\begin{equation}
\mathcal{A}
=\sum_{k=0}^{N}
\frac{\binom Nk^2}{\binom{2N}{2k}}
\tilde r^{\,4kN_T},
\qquad
\tilde r=1-2\lambda,
\label{eq:Amg}
\end{equation}

which reduces to

\[
\mathcal{A}\longrightarrow\sqrt{\pi N}
\]

in the noiseless limit by Stirling's approximation. A pure matchgate circuit therefore has an IPR that grows as $\sqrt{N}$ rather than approaching the Porter--Thomas value.

SWAP gates preserve the Majorana basis but redistribute weight between neighbouring degree sectors. Averaging over uniformly distributed monomials within each sector gives an exact tridiagonal transfer matrix acting on the vector of sector weights, whose derivation is given in Appendix~\ref{app:transfer}. The sectors $k=0$ and $k=N$, corresponding to the identity and global fermion parity, remain invariant, and in the absence of noise the stationary value is

\begin{equation}
\mathcal{A}_{\mathrm {eq}}
= 2+\frac{(2^N-2)^2}{2^{2N-1}-2}
\longrightarrow4,
\label{eq:parity}
\end{equation}

with the limit equal to $4$ because the dynamics are confined to a single fermion-parity sector of dimension $D/2$.

The complete noisy evolution is therefore described by

\begin{equation}
W
\longmapsto
\Delta(\tilde r)\,M^{n_S}\,W,
\qquad
\Delta=\mathrm{diag}\!\left(\tilde r^{4k}\right),
\label{eq:opdegree}
\end{equation}

where $M$ is the SWAP transfer matrix and $n_S$ is the number of SWAP gates in one circuit layer. Iterating this $(N+1)\times(N+1)$ map gives the exact Majorana degree distribution for any system size, from which the truncation order $k^*$ follows directly. The competition is between SWAP gates, which move weight towards higher degree, and dephasing, which suppresses the higher-degree sectors. This defines the Majorana-propagation boundary of Fig.~\ref{fig:extabstractfigure}(b).

\subsection{Moments Do Not Determine Truncation Cost}

Two features of this family expose a mismatch that is independent of noise. First, for a pure matchgate circuit, the sector weights remain fixed at $W_{2k}=\binom Nk$, so the discarded weight of Eq.~\eqref{eq:degree_tail} becomes small only when the truncation order $k^*$ is of order $N$. Majorana propagation therefore fails, even though the circuit is solved in $O(N^3)$ time by the covariance-matrix algorithm. The two fermionic algorithms exploit different structures: one uses Gaussianity, and the other relies on low Majorana degree. The second is obtained by comparing truncations in different operator bases. A degree-two Majorana monomial $\gamma_a\gamma_b$ carries a Jordan--Wigner string whose qubit weight is proportional to the separation of $a$ and $b$, so its average Pauli weight grows linearly with $N$. A Pauli-weight truncation therefore discards precisely the operators that a Gaussian circuit conserves, whereas a Majorana-degree truncation retains them. The same circuit therefore appears easy in one basis and difficult in the other.

The common principle is that truncation algorithms are governed by the distribution of operator weight across the basis adapted to the dynamics. Their cost is determined by the smallest cutoff for which the discarded weight falls below the target accuracy, whereas diagnostics such as the IPR and purity depend only on low-order moments of the same distribution. The matchgate example therefore shows that a moment cannot, in general, determine the complexity of a truncation algorithm.

\subsection{Non-Gaussianity}

The Majorana sector description cannot detect fermionic non-Gaussianity. It retains only the total weight within each degree sector, whereas Gaussianity is determined by the correlations between Majorana monomials through Wick's theorem. The non-Gaussianity measure $\delta$ of Eq.~\eqref{eq:nongaussianity} is therefore evaluated by exact density-matrix simulation of the brickwork circuit, averaged over circuit realisations, restricting the calculation to $N\le8$.

The measure is well defined under the present noise model. Matchgates preserve Gaussianity, and Majorana dephasing preserves every Wick relation, so dephasing alone cannot generate non-Gaussianity. Numerically, $\delta$ vanishes to machine precision without SWAP gates and increases monotonically with circuit depth once SWAP doping is introduced.

Fig.~\ref{fig:extabstractfigure}(b) shows the resulting non-Gaussianity boundary for two numerical resolutions. At the experimentally motivated resolution, the boundary crosses the uncovered region exactly as in the Clifford map, demonstrating that the mismatch persists even when the resource is the physically relevant one. A finer resolution shifts the boundary but does not alter this conclusion.

\section{Discussion and outlook}
\label{sec:discussion&outlook}

Through this work, we observe that local  noise separates dynamical diagnostics from classical simulation through a simple and general mechanism. Classical simulation thresholds depend on second-order properties of the noisy output state, whereas the dynamical resources considered here, mixed-state magic, scrambling, and fermionic non-Gaussianity, depend on fourth-order correlations. Because local depolarising noise suppresses higher-order moments faster than second-order moments, these two classes of quantities evolve on different noise and depth scales. This difference in moment order arises because an approximation error of the output distribution is quadratic in the state by construction, while the Clifford group is a unitary $3$-design, so that magic is invisible below the fourth moment. More generally, whenever the free operations of a resource theory form a $k$-design, the resource first appears at the $(k+1)$-th moment while the error that governs simulation remains second order, and some version of the separation we describe should be expected.

For Clifford circuits, this separation is demonstrated exactly using the fourth-order transfer-network formalism, which yields finite-size boundaries for mixed-state magic, scrambling, and anticoncentration. The resulting phase diagrams show that dynamical resources becoming trivial does not coincide with the onset of efficient classical simulation. An intermediate regime is observed in which the output state is efficiently simulable while still exhibiting nontrivial higher-order structure.

The same phenomenon persists in matchgate circuits. Although the physical resource changes from mixed-state magic to fermionic non-Gaussianity, the classical simulation threshold is again controlled by a second-order quantity through fermionic degree truncation. The agreement between the two circuit families suggests that the mismatch is not architecture-specific, but rather a consequence of the different moment orders governing dynamical diagnostics and state-dependent simulation algorithms.

Our results therefore clarify the relationship between dynamical signatures of quantum complexity and efficient classical simulation.  A measured value of magic or of an out-of-time-order correlator, on its own, therefore does not carry much implication about simulation cost under noise. This motivates the search for dynamical quantities whose behaviour under noise is tied more closely to the second-order structure that most known algorithms actually consume. 

Several directions follow. From a complexity-theoretic perspective, the regions that are not ruled out by the union of classical algorithms carry no hardness guarantee. Establishing an unconditional lower bound for efficient classical sampling would amount to resolving questions closely related to the separation of BPP and BQP, which remains open. A more realistic objective is to derive lower bounds within restricted families of simulation algorithms, such as Pauli-path or Majorana-degree truncation methods, thereby distinguishing the limitations of particular algorithmic approaches from unconditional classical hardness. It would also be interesting to place these diagnostics within the emerging partition-function formulation of quantum resources~\cite{salazar2026stabilizerstatisticalmechanicsframework}, in which the Pauli spectrum is encoded in a stabilizer partition function whose free energy generates a family of magic monotones~\cite{salazar2026stabilizerstatisticalmechanicsframework}. Whether this resource-theoretic partition function connects to the statistical-mechanical partition functions whose evaluation is $\#\mathrm{P}$-hard, and which underlie the hardness arguments for quantum sampling, is open. Finally, our results are reminiscent of the notion of pseudomagic, where ensembles with low nonstabilizerness are computationally indistinguishable from ensembles with high nonstabilizerness \cite{Gu_2024}. It is shown that low-order measurements can be fooled by cryptographic means: what we find is that noise achieves something of the same kind physically and without any assumption, by moving the resource and the cost onto different scales. Both echo a similar underlying tension: a quantity that can be estimated from a realistic number of copies is a low-degree polynomial in the state, whereas simulation cost is a property of the whole distribution, and it is not clear that any efficiently measurable quantity can be faithful to it. Exploring that question would determine what a noisy device can honestly report about its own hardness.

\section*{Acknowledgements}
Strelchuk acknowledges support from the Wellcome Leap as part of the Q4Bio Program and the Royal Society University Research Fellowship. Subramanian acknowledges support from the Royal Society through a University Research Fellowship.

\section*{AI Disclosure}
Generative AI assistants were used to aid  literature discovery, code refinement, figure styling, and consistency checks of intermediate analytical and numerical calculations.

\printbibliography

\appendix

\section{Gate-set orthogonality and Pauli paths}\label{app:orthogonality}
 
The Pauli-path bound (Eq.~\eqref{eq:pauli_path}) requires that distinct paths be uncorrelated in expectation. We reproduce the argument of Ref.~\cite{Aharonov_2023} in the form used here, and check that it survives the doping.
 
\paragraph{Right invariance.} A distribution $\mathcal{D}$ on unitaries is right-invariant if $\mathbb{E}_{U\sim\mathcal{D}}[F(U)]=\mathbb{E}_{U\sim\mathcal{D}}\mathbb{E}_{V\sim \mathcal{P}}[F(UV)]$ for every $F$. The uniform distribution on the Clifford group has this property. $CV$ is Clifford for every Clifford $C$ and Pauli $V$, and for fixed $C'$ the equation $CV=C'$ has exactly $4^N$ solutions, one per Pauli, so reindexing the double sum by $C'=CV$ returns the original average. The same holds for a uniformly random two-qubit Clifford on each bond.
 
\paragraph{Gate-set orthogonality.} For $P\neq Q$ in $\mathcal{P}$ and any right-invariant $\mathcal{D}$,
\begin{equation}
\mathbb{E}_{U\sim\mathcal{D}}\left[UPU^\dagger\otimes UQU^\dagger\right]=0 .
\label{eq:orthogonality}
\end{equation}
Indeed, inserting a random Pauli and using $VPV^\dagger=(-1)^{\langle V,P\rangle}P$, where $\langle V,P\rangle$ is $1$ if $V$ and $P$ anticommute and $0$ otherwise,
\begin{align}
\mathbb{E}_{V}\left[VPV^\dagger\otimes VQV^\dagger\right]
&=\frac{1}{4^N}\sum_V(-1)^{\langle V,P\rangle+\langle V,Q\rangle}\,P\otimes Q\notag\\
&=\frac{1}{4^N}\sum_V(-1)^{\langle V,PQ\rangle}\,P\otimes Q=0,
\end{align}
because $PQ\neq\mathbb{I}$ up to phase, and a nonidentity Pauli commutes with exactly half of the Pauli group. Conjugating by $U$ and averaging gives Eq.~\eqref{eq:orthogonality}.
 
\paragraph{Orthogonality of Fourier paths.} For two distinct paths $s\neq s'$, the Fourier coefficients factorise over layers, so
\begin{equation}
\mathbb{E}_{C\sim\mathcal{D}}\left[f(C,s,x)f(C,s',x)\right]
=\beta\prod_{i}\mathbb{E}_{U_i}\!\left[\cdot\right],
\end{equation}
with $\beta$ collecting the boundary overlaps. Let $i$ be the first layer at which the two paths differ. Its contribution is, by cyclicity of the trace,
\begin{equation}
\tr\!\left[(s_{i+1}\otimes s'_{i+1})\,\mathbb{E}_{U_i}\!\left[(U_i\otimes U_i)(\tilde s_i\otimes \tilde s'_i)(U_i^\dagger\otimes U_i^\dagger)\right]\right],
\end{equation}
where $\tilde s_i=Ks_iK^\dagger$ and $\tilde s'_i=Ks'_iK^\dagger$ account for the fixed non-Clifford gates $K$ of that layer. Since $K$ is fixed and $s_i\neq s'_i$, the two conjugated strings are distinct Pauli strings up to phase, so the inner expectation vanishes by Eq.~\eqref{eq:orthogonality} and the whole average is zero. The doping therefore does not spoil the hypothesis, for the global and for the brickwork architecture alike.

\section{Numerics and Code}
\label{app:numerics}
The code used to reproduce the analytical and numerical results and figures in this work is available at \url{https://github.com/anjaliwgh/noisy-quantum-diagnostics-simulability}.

\subsection{Numerical Results}

\paragraph{Sampling convergence.}
The non-Gaussianity of Fig.~\ref{fig:extabstractfigure}(b) is averaged over 20 circuit
realisations at $N=6$.

\paragraph{Ratio of decay rates.} The factor of two between the decay rates of the fourth and second moments in Eq.~\eqref{eq:ratio} is derived for
the undoped global model. On the doped brickwork, where this derivation does not apply, the measured ratio remains $2.00$--$2.01$ for sparse noise over $p=0.02$--$0.10$ and equals $2.00$ for dense noise at weak depolarisation, supporting the mechanism proposed in Sec.~\ref{sec:mechanism}. The decay rate of the excess inverse participation ratio is independent of system size to five or six significant digits for $N=10$--$12$.

\paragraph{Truncation order.} The Chernoff upper bound on the Pauli-path truncation order is independent of system size under dense noise, in agreement with Eq.~\eqref{eq:ellstar}. Both the upper and lower bounds grow linearly with $N$ under sparse noise. This shows that the linear scaling is a property of the truncation order itself. 

\paragraph{Purity threshold and crossover.}
The exact depth at which the purity reaches $D^{-1/3}$ grows linearly with system size up to $N=22$ and remains within a few layers of the analytical bound of Eq.~\eqref{eq:entropy}. In contrast, the uniform-sampling threshold is independent of $N$ from $N=8$ onward. The depth at which the magic witness changes sign closely follows the purity threshold once the injected magic has saturated, validating the use of purity as its large-system proxy. Under dense noise, the witness remains non-positive for all depths examined.

\subsection{Method Validation}

\paragraph{Commutant and one-layer transfer map.}
The Gram matrix has rank 30 for $N\ge3$ and rank 29
at $N=2$. Since every single-site Gram
entry is a power of two and every entry of the $T$-map is an integer, the layer map is a
rational matrix in $D=2^N$, and its block structure and the closed form of the $\Omega_4$
block can be verified; these hold exactly for $N=3,\ldots,12$.

\paragraph{Independent verification.}
The fourth-moment network, the replica transfer matrix, and the Pauli-path construction are implemented independently. Their predictions for the inverse participation ratio agree, including in the presence of doping. Comparison with explicit random two-qubit Clifford circuits at $N=4$ further reproduces the purity, stabilizer purity, and collision probability.

\paragraph{Witness validation.}
For circuits composed solely of Clifford gates and depolarising noise, the mixed-state magic witness satisfies $\mathcal{W}\le0$ identically. Across all numerical realizations, the maximum observed value is consistent with zero, confirming that the implementation does not produce false positive witnesses.

\paragraph{Matchgate sector dynamics.}
The SWAP transfer matrix agrees with brute-force conjugation of Majorana monomials at
$N=4$, the asymptotic form of Eq.~\eqref{eq:Amg} is reproduced to a relative $10^{-4}$ at
$N=1024$, and iterating the SWAP walk reproduces Eq.~\eqref{eq:parity} to eight digits.
Against exact simulation at $N=6$, the sector formula is accurate once the depth is
comparable with the system size and overestimates the concentration of the output at
shallower depth, as expected from the equilibration assumption.

\section{The commutant and the T gate map}
\label{app:commutant}

\paragraph{The anticommuting-pair identity.} 
For a pure state, 
\begin{equation}
    R\rho R=D^{-1}\sum_Pa_P(-1)^{\langle R,P\rangle}P
\end{equation} gives \begin{equation}
    \mathrm{Tr}(R\rho R\rho)=D^{-1}\sum_Pa_P^2(-1)^{\langle R,P\rangle},
\end{equation} whose left-hand side is $a_R^2$. Multiplying by $a_R^2$ and summing over $R$ yields 
\begin{equation}
    \sum_{P,Q}(-1)^{\langle P,Q\rangle}a_P^2a_Q^2=D^2\Delta_4
\end{equation}
and combining with $\sum_{P,Q}a_P^2a_Q^2=D^2$ separates the ordered pairs into $\sum_{\{P,Q\}=0}a_P^2a_Q^2=\tfrac{D^2}{2}(1-\Delta_4)$. Summing Eq.~\eqref{eq:uv} over the $D^2/4$ pairs after a Clifford twirl and using this identity gives the map leading to Eq.~\eqref{eq:delta4t}.

\section{Transfer matrices}
\label{app:transfer}

\paragraph{Global second moments.}
The second moments of the global Clifford model are generated by a $2\times2$ transfer matrix acting on the $k=2$ commutant. Defining
\[
\widetilde G=
\begin{pmatrix}
D^2 & D\\
D & \Lambda_2^N
\end{pmatrix},
\]
with $G$ the corresponding Gram matrix, the transfer matrix is
\[
B=G^{-1}\widetilde G=
\begin{pmatrix}
1 & \beta\\
0 & \kappa_2
\end{pmatrix},
\]
where
\[
\beta=\frac{D^2-\Lambda_2^N}{D(D^2-1)}.
\]
The eigenvalue $1$ reflects trace preservation, while $\kappa_2$ is the damping factor defined in Eq.~\eqref{eq:kappa}. The relation
\[
\frac{\beta}{1-\kappa_2}=\frac1D
\]
ensures that repeated application of $B$ converges to the uniform state. Contracting $B^{N_T-1}$ with the initial vector reproduces the closed-form expressions of Eq.~\eqref{eq:closed2}. 

\paragraph{Brickwork second moments.}

For the brickwork architecture, the transfer matrix acts on replica labels in $\{e,s\}^N$. A two-qubit Clifford gate satisfies
\[
T(e|ee)=T(s|ss)=1,\qquad
T(e|es)=T(s|es)=\frac{d}{d^2+1},
\]
Depolarising noise on each noisy qubit is represented by
\[
\begin{pmatrix}
1 & (1-r^2)/d\\
0 & r^2
\end{pmatrix}.
\]
The purity is obtained by contracting the network with boundary weights $d$ for the $e$ replica and $d^2$ for the $s$ replica, and the normalised IPR uses weight $d$ for both replicas.

\paragraph{SWAP transfer matrix.}

A nearest-neighbour SWAP is supported on the four Majoranas of the exchanged modes and maps every Majorana monomial to a single monomial. On the four-mode block, the local occupancy satisfies

\[
m\longmapsto4-m,
\qquad
m=1,2,3,
\]

while the identity and local parity are unchanged. Consequently, the total Majorana degree changes only by $\pm2$ or $0$.

Under the assumption of uniform weight within each degree sector, the occupancy of a random monomial of degree $2k$ follows the hypergeometric distribution

\[
P(m|2k)=
\frac{\binom4m\binom{2N-4}{2k-m}}
{\binom{2N}{2k}},
\]

which yields the tridiagonal, column-stochastic transfer matrix

\begin{align}
M_{k\pm1,k} &= \frac{4\binom{2N-4}{2k-2\pm1}}
{\binom{2N}{2k}},\\
M_{k,k} &= \frac{ \binom{2N-4}{2k} +6\binom{2N-4}{2k-2} +\binom{2N-4}{2k-4}} {\binom{2N}{2k}}.
\end{align}

The matrix is symmetric under $k\mapsto N-k$, and the sectors $k=0$ and $k=N$ are absorbing, corresponding to the identity and global fermion parity.

\end{document}